\documentclass[twocolumn,pra]{revtex4}

\usepackage{graphicx}
\usepackage{dcolumn}
\usepackage{bm}

\newcommand{\beq}{\begin{equation}}
\newcommand{\eeq}{\end{equation}}
\newcommand{\beqa}{\begin{eqnarray}}
\newcommand{\eeqa}{\end{eqnarray}}
\newcommand{\beqar}{\begin{eqnarray*}}
\newcommand{\eeqar}{\end{eqnarray*}}

\def \la {\langle}
\def \ra {\rangle}

\begin{document}
\input epsf
\title{\bf \large On the efficiency of nonlocal gates generation}

\author{Berry Groisman$^a$ and Benni Reznik$^b$}
\affiliation{ $^a$HH
Wills Physics Laboratory, University of Bristol, Tyndall Avenue,
Bristol, BS8 1TL, UK
\\ $^b$  School of Physics and Astronomy,
Raymond and Beverly Sackler Faculty of Exact Sciences, Tel Aviv
University, Tel Aviv 69978, Israel}


\begin{abstract}
We propose and study a method for using non-maximally entangled
states to implement probabilistically non-local gates. Unlike
distillation-based protocols, this method does not generate a
maximally entangled state at intermediate stages of the process.
As a consequences, the method becomes more efficient at a certain
range of parameters. Gates of the form
$\exp[i\xi\sigma_{n_A}\sigma_{n_B}]$ with $\xi\ll1$, can be
implemented with nearly unit probability and with vanishingly
small entanglement, while for the distillation-based method the
gate is produced with a vanishing success probability. We also
derive an upper bound to the optimal success probability and show
that in the small entanglement limit, the bound is tight.

\end{abstract}


\maketitle

\section{Introduction}\label{intro}

Over the recent years it has been shown that entanglement can be
used to perform various quantum process such as teleportation,
quantum communication, quantum cryptography and quantum
computation \cite{nielsen-book}. One important use of
entanglement, involves the implementation of quantum gates (or
more generally, interaction) between spatially separated qubits,
without actually having to transport the physical system that
carries the state from one place to another. The study of such
``non-local" gate operations, has some bearing on practical and
fundamental issues. On one hand, the connection between
entanglement as a resource for non-local gates, and the related
problem of quantifying the capability of non-local gates to
generate entanglement, deals with the fundamental relation between
entanglement and interactions. This problem has been studied by
several groups
\cite{nlham,plenio,krauscirac,popescu,huelga,stator,vidal+cirac,nlPOVM}
but is yet not fully understood. On the other hand, non-local
gates may be used as primitives in protocols involving several
separated systems, for instance, in multi-party computation
problems.

In principle, any non-local gate can be implemented given by
sufficient amount of shared entanglement, and by exchanging a
sufficient number of classical bits: we can use quantum
teleportation to teleport the relevant qubits states to a single
location, apply {\em locally} the relevant interaction in order to
generate the desired gate, and finally teleport the states back to
their original position. In general however, the implementation of
gates with quantum teleportation methods may not be efficient, and
provides only an upper bound on the required amount of
entanglement. Since entanglement is an expensive resource, it is
important to optimize its usage and search for economic methods
for implementing nonlocal gates.

In this paper we shall reconsider the problem of gates acting on
two qubits which have the structure
 \beq\label{gate}
U_{AB}(\xi)=e^{i\xi\sigma_{n_A}\sigma_{n_B}}. \eeq
  This family of gates includes the controlled-NOT (CNOT) gate, which up to
   local rotations corresponds to $U_{AB}(\xi=\pi/4)$.
By using teleportation, $U_{AB}(\xi)$ can  be implemented for any $\xi$
 using two maximally entangled pairs (e-bits). It was shown
however that, if $n_A$ and $n_B$ are locally known, one e-bit is
sufficient \cite{huelga,stator}. Moreover, it was recently shown
that for the cases $\xi=\pi/2^N$, where $N>2$, this gate can be
implemented with
 less than one e-bit \cite{vidal+cirac}. The later method
  utilizes $N$ pairs, each carrying a different amount of
entanglement, that sums up to less than one e-bit.

In this article we shall study the possibility of realizing the
gate (1) using a {\em single} copy of a {\em non-maximally} entangled
state
 \beq\label{ebit}
|\phi_{ab}({\alpha})\ra=\cos\alpha|0_a\ra|0_b\ra+\sin\alpha|1_a\ra|1_b\ra,
\eeq \noindent In this case, as we show in the sequel, with
certain states the gate can not be implemented deterministically
\cite{open}. Hence in the rest of the paper we shall study methods
for probabilistic implementations.

\begin{figure} \epsfxsize=3.truein
      \centerline{\epsffile{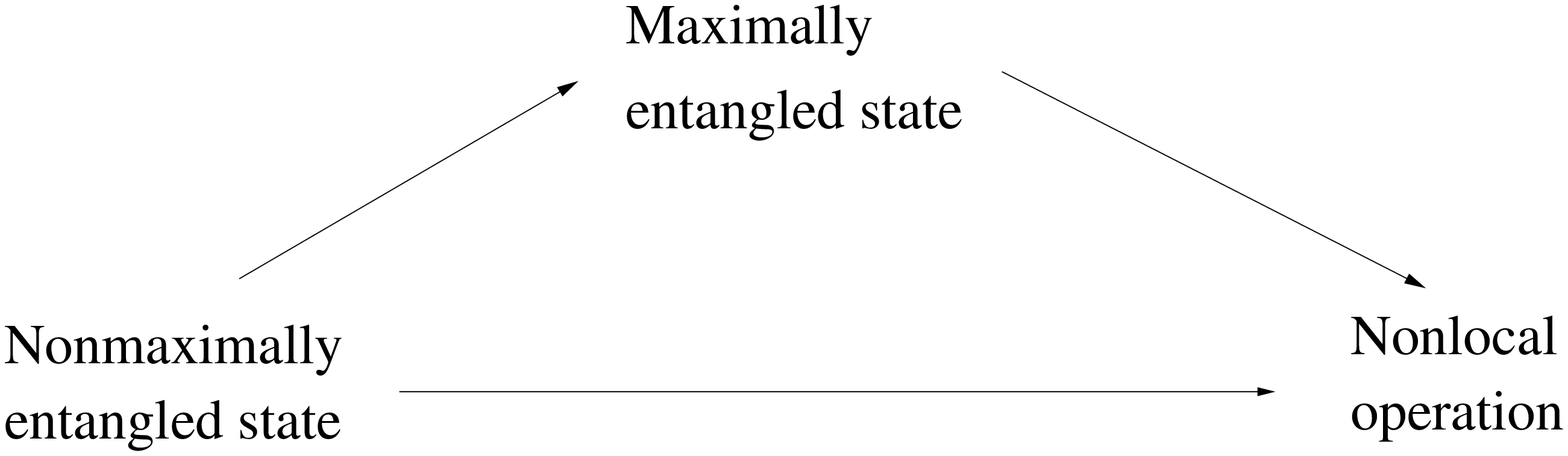}}
  \caption[]{A non-local gate can be implemented using
   distillation-based method which starts by generating a maximally
   entangled state in an intermediate stage of the protocol, or
   by a more direct method that avoids distillation.
  }\label{scheme} \end{figure}

The simplest method to generate (1) is offered by distillation
methods. To this end, one can first transform with some
probability the state (\ref{ebit}) to a the maximally entangled
pair \beq\label{mebit}
|\phi_{ab}^+\ra=\frac{1}{\sqrt{2}}(|0_a\ra|0_b\ra+|1_a\ra|1_b\ra),
\eeq

\noindent and then utilizes a deterministic scheme (e.g. that
suggested in \cite{stator} in order to implement the gate (see
Fig. (\ref{scheme})).
 In this scheme, the  probability to implement the gate successfully
  is determined by the probability,
  $p_{dis}$ to convert the non-maximally entangled pair
(\ref{ebit}) into maximally entangled pair (\ref{mebit}).
It was recently shown that the maximal value of $p_{dis}$
is twice the modulus square of the Schmidt coefficient of
smaller magnitude \cite{lopopescu, vidal}, i.e.
$p_{dis}=2(\sin\alpha)^2$ and $0<\alpha<\pi/4$.
The simple method to distill an e-bit out of a non-maximally
entangled state is known as the  Procrustean method \cite{proc_meth}.

We notice that the success probability of a distillation-based
method does not depend on the gate parameter $\xi$, and decreases
with $\alpha$. This suggests that in a more efficient method one
could optimize the scheme as a function of the gate parameter
$\xi$ as well. As we show, an improvement of the
distillation-based protocols can be indeed obtained when the
process does not require in an intermediate step a maximally
entangled state (Fig. 1).

To see that this is possible, consider the
special case of $\alpha=\xi$.
Let Alice and Bob start by
performing local CNOT interactions: Alice applies CNOT (with
respect to $\sigma_{n_A}$) between her qubits $a$ and
$A$ , described by the unitary transformation \beq\label{cnot}
U_{aA} = |0_a\ra\la 0_a| \otimes I_{A} + |1_a\ra\la1_a|\otimes
\sigma_{n_A}. \eeq Bob does the same on his side. This yields the
four-particle state \beq
\label{aftercnot}\biggl(\cos\alpha|0_a0_b\ra\otimes I_{AB}
 + \sin\alpha|1_a1_b\ra \otimes\sigma_{n_A}\sigma_{n_B} \biggr)|\Psi_{AB}\ra
\eeq

Next they measure $a$ and $b$: Alice measures $\sigma_y$ and  Bob
measures $\sigma_x$. The result can be then written as \beq
\biggl(\cos\alpha~ I_{AB} \pm i  \sin\alpha~
\sigma_{n_A}\sigma_{n_B} \biggr) |\Psi_{AB}\ra= e^{\pm
i\alpha\sigma_{n_A}\sigma_{n_B}}|\Psi_{AB}\ra , \eeq where the $+$
($-$) sign is obtained with probability 1/2 when the measurement
results satisfies $\sigma_x=\sigma_y$ ($\sigma_x=-\sigma_y$).
Hence we showed that the success probability for gates with
$\alpha=\xi$ is here given by 1/2! which can be larger compared
with $p_{dis}$ for small enough values of $\alpha$.

At first sight this result seems paradoxical \cite{thanks-sandu}.
Recall the entanglement capability the gate (1) is known in this
limit to be $\sim \xi$ \cite{nlham}. Hence we have started with an
entanglement proportional to $\xi^2=\alpha^2$ and ended up with
more entanglement $\sim \xi$ with probability 1/2! The resolution
to this paradox, is connected with the fact that the maximally
capability is obtained by acting the gate on a partially entangled
state. Hence in 1/2 of the times we end up by {\em reducing} the
entanglement. The net increase is not larger then $\xi^2$.

In the rest of this paper we present and study a method for
generating gates, that does not involve maximal entanglement
states at intermediate steps, and hence utilizes the above
mentioned idea of a ``direct" map between entanglement and gates
(Fig. 1). The method can be used for arbitrary values of $\xi$ and
$\alpha$, and for a certain range of parameters, is more efficient
compared with the distillation-based (e.g. Procrustean) methods.
In particular, we show that gates with very a small angle $\xi$,
can be implemented with probability very close to unity, with
vanishingly small entanglement. This is to be compared with the
distillation-based method, which in this case succeeds with a
vanishingly small probability.

We shall also exam the general relation between the maximal
success probability $p_{max}$, the gate parameter $\xi$, and the
given entangled state, described by the parameter $\alpha$.


The article proceeds as follows. In the next section
\ref{secbound} we derive a bound on $p_{max}$ for a certain range
of parameters $\xi$ and $\alpha$. In Sec. \ref{method} we present
a ``direct"  method for probabilistic generation of non-local
gates. Then, in Sec. \ref{optgate} we refer to one of these gates
as a target gate and analyze the probability of its
implementation. Based on these results we show that the bound
derived in Sec. \ref{secbound} is tight.

\section{A Bound on the success probability}\label{secbound}
In this section we derive an upper bound on the maximal
probability of successful implementation of the gate (\ref{gate})
on the state $|\Psi_{AB}\ra$ using (\ref{ebit}) as a resource when
$\alpha<\xi$.

We can write a required task as a transformation \beqa
(\cos\alpha|0_a\ra|0_b\ra+\sin\alpha|1_a\ra|1_b\ra)|\Psi_{AB}\ra\longrightarrow^p\nonumber\\
e^{i\xi\sigma_{n_A}\sigma_{n_B}} |\Psi_{AB}\ra
=(\cos\xi+i\sin\xi\sigma_{n_A}\sigma_{n_B})|\Psi_{AB}\ra, \eeqa

\noindent where the $p$ denotes the probability of the above
transition.

Let us consider the special case of the product state

$$|\Psi_{AB}\ra=|\uparrow_{n^{\bot}_A}\ra|\uparrow_{n^{\bot}_B}\ra,$$
where the axis $n^{\bot}$ are such that
$\sigma_n|\uparrow_{n^{\bot}}\ra=|\downarrow_{n^{\bot}}\ra$. In
this case

\beqa\label{majo2}
(\cos\alpha|0_a\ra|0_b\ra+\sin\alpha|1_a\ra|1_b\ra)|\uparrow_{n_A}\ra|\uparrow_{n_B}\ra\longrightarrow^p\nonumber\\
\cos\xi|\uparrow_{n_A}\ra|\uparrow_{n_B}\ra+i\sin\xi|\downarrow_{n_A}\ra|\downarrow_{n_B}\ra
\eeqa

The main idea is to use the {\it majorization condition}
\cite{lopopescu, nielsen, vidal} in order to bound the probability
$p$. If $\alpha<\xi$, then $\{(\cos\alpha)^2,(\sin\alpha)^2\}$
does not majorized by $\{(\cos\xi)^2,(\sin\xi)^2\}$, which means
that the transformation (\ref{majo2}) cannot be performed with
certainty. The maximal probability of this transformation given in
this case by

\beq\label{bound} p_{max}=\frac{(\sin\alpha)^2}{(\sin\xi)^2}. \eeq

Now we apply to the general result from \cite{U-U} that shows that
the probability of successful simulation of a unitary is
independent of the input state. Thus, the bound (\ref{bound})
obtained for the product state is as good as a bound one might
obtain for all other $|\Psi_{AB}\ra$. In our particular case it
can be also easily checked explicitly that any other
$|\Psi_{AB}\ra$ will give the same upper bound on $p$.

In the Sec. \ref{optgate} we will show that for $\alpha\ll\xi$
this bound is tight.

In passing we note that for $\xi=\pi/4$ the bound (\ref{bound}) is
consistent with the results of \cite{U-U,classes} concerning an
optimal simulation of non-local CNOT gate, i.e.
$U_{AB}(\xi=\pi/4)$, using $U_{AB}(\xi<\pi/4)$.

We note that for $\alpha\geq\xi$ the majorization method does not
provide any restrictions on the transformation (\ref{majo2}), i.e.
we cannot use it to derive the bound in this case. The
majorization
 does not prevent $p_{max}$ even from reaching one
  when $\alpha=\xi$, although for small $\xi$ this possibility
 can be discarded based on the results on the entanglement
 capability \cite{nlham}. It is natural to conjecture that
 in the single state case,
 $p_{max}=1$ may be achieved only for $\alpha=\pi/4$.
 It is unlikely that $p_{max}$
 exhibits non-smooth behavior. Nevertheless,
 the question whether the gate (\ref{gate}) can
 be generated deterministically for $\alpha\geq\xi$ is still open.

\section{A method for probabilistic generation of non-local gates}\label{method}
In this section we demonstrate how Alice and Bob can
probabilistically generate a pair of gates of the type
(\ref{gate}) directly from $|\phi_{ab}({\alpha})\ra$  without
distilling it.

\subsection{Mapping states to Stator}
In order to simplify the explanation of our method we use a hybrid
state-operator object (stator) defined in \cite{stator}. The
Stator describes quantum correlations between the state of one
systems and the operation/s acting on another system/s.

How do we prepare the stator? Alice and Bob start by performing
the first step as described in Sec. \ref{intro} - Eq. (\ref{cnot})
and (\ref{aftercnot}).

 Next Bob performs a measurement of $\sigma_x$ of the
 entangled qubit $b$ to project out a certain value.   The resulting state is now
\beq\label{resx} \bigl(|0_b\ra \pm |1_b\ra\bigr)\otimes
\biggl(\cos\alpha|0_a\ra\otimes I_B \pm \sin\alpha|1_a\ra\otimes
\sigma_{n_A}\sigma_{n_B} \biggr) |\Psi_{AB}\ra \eeq Finally Bob
informs Alice what was the result of his measurement
 by sending Alice one classical bit of information.
For the case that $\sigma_x=-1$ Alice performs a trivial $\pi$
rotation around the $\hat z$ axis and flips the $-$ sign to a $+$
sign. The resulting state of the system is now given by
\beq\label{Spsi} \bigl(|0_b\ra \pm |1_b\ra\bigr)\otimes
\biggl(\cos\alpha|0_a\ra\otimes I_{B} + \sin\alpha|1_a\ra\otimes
\sigma_{n_A}\sigma_{n_B} \biggr) |\Psi_{AB}\ra \eeq

Since Bob's previously entangled qubit factors out,
the final state of Alice's qubits $A$, $a$ and Bob's qubit $B$ is

\beq\biggl(\cos\alpha|0_a\ra\otimes I_{B} +
\sin\alpha|1_a\ra\otimes\sigma_{n_A}\sigma_{n_B} \biggr)
|\Psi_{AB}\ra\equiv S|\Psi_{AB}\ra,\eeq

\noindent where the stator $S$ captures the correlation between
the state $|0_a\ra$ and $|1_a\ra$ of Alice's qubit $(a)$ and
unitary transformation $I_{AB}$ and $\sigma_{n_A}\sigma_{n_B}$
acting on $|\Psi_{AB}\ra$.

If $\alpha=\pi/4$, then the stator is ``maximal" and can be used
to apply (\ref{gate}) with certainty \cite{stator}. This is done
by utilizing the identity:
 \beq e^{i\xi \sigma_{x_a}} S = S
e^{i \xi \sigma_{n_A}\sigma_{n_B}} .
  \eeq
  Hence by applying a
local rotation by Alice we can ``pull out" the required unitary
operator that generates the gate. In the following section we show
how to use ``nonmaximal" stator in order to apply (\ref{gate})
probabilistically.

\subsection{Generating non-local gates}
Our goal is to find an appropriate $\theta(\xi,\alpha)$ such that
by applying locally a unitary on $S$ one gets the sum of
probabilistic unitaries

\begin{widetext}
\beq\label{2U} e^{i\theta(\xi,\alpha)\sigma_{x_a} } S=
\sqrt{p}~|0_a\ra \otimes e^{i\xi \sigma_{n_A}\sigma_{n_B}}
+\sqrt{1-p} ~|1_a\ra \otimes
\sigma_{n_A}\sigma_{n_B}e^{i\tilde{\xi} \sigma_{n_A}\sigma_{n_B}}.
\eeq \end{widetext}

\noindent As a result the state of Alice's particle $a$ will be
correlated with two different gates applied on the state
$|\Psi\ra_{AB}$: if Alice will measure her particle $a$ in
$z$-basis, then she will get $|0_a\ra$ with probability $p$ and
the nonlocal gate
$U_{AB}^{(1)}=\exp(i\xi\sigma_{n_A}\sigma_{n_B})$ will be
generated. Similarly, when she gets $|1_a\ra$ (that happens with
probability $1-p$) then
 the
 nonlocal gate
$U_{AB}^{(2)}=\exp(i\tilde{\xi}\sigma_{n_A}\sigma_{n_B})$
is generated (up to trivial local rotations $\sigma_{n_B}\sigma_{n_B}$).\\

 It is straightforward  to verify that Alice has to apply in
(\ref{2U}) the unitary with

\beq\label{theta}\label{tg2xitheta}
\theta(\xi,\alpha)=\tan^{-1}\biggl[\frac{\tan\xi}{\tan\alpha}\biggr]
\eeq
\noindent and to show that $\xi$, $\tilde{\xi}$, $p$, and $\alpha$
have to satisfy the following relations:

\beq p=(\cos\alpha \cos{\theta})^2+(\sin\alpha\sin{\theta})^2,\eeq
\beq\label{xi1theta} \xi=\cos^{-1}\biggl[\frac{\cos\alpha
\cos{\theta}}{\sqrt{p}}\biggr] \eeq

\beq\label{xi2theta} \tilde{\xi}=\cos^{-1}\biggl[\frac{\sin\alpha
\cos{\theta}}{\sqrt{1-p}}\biggr] \eeq

\beq\label{xixi} \frac{\tan{\xi}}{
\tan{\tilde{\xi}}}=(\tan\alpha)^2 \eeq

\beq\label{p}
p=\frac{(\sin\alpha)^2}{1-\frac{1-2(\sin\alpha)^2}{1-(\sin\alpha)^2}\cos^2{\xi}}
\eeq


Thus, as it follows from (\ref{xixi}), given a nonmaximally
entangled pair (\ref{ebit}) as a resource Alice and Bob can
generate any the pair of gates  $U_{AB}^{(1)},U_{AB}^{(2)}$ (each
one with appropriate probability)
 where two parameters $\xi,
\tilde{\xi}$ which satisfy (\ref{xixi}). Since for fixed $\alpha$
infinitely many pairs of angles
 $\xi,\tilde{\xi}$ (in the range $\pm \pi/2$) which satisfy (\ref{xixi}) can be found, infinitely many
pairs of gates can be generated.\\
To obtain the required gate they choose a pair $\xi,\tilde{\xi}$ that
satisfies (\ref{xixi}), and calculate $\theta$ according to
(\ref{tg2xitheta}). Finally Alice applies
  $U_a=\exp(i\theta\sigma_{x_a})$, and  measures the operator $\sigma_z$ of
qubit $a$.

 We note that for the special case of "maximal" $S$,
i.e. when $\alpha=\pi/4$, we obtain $p=\frac{1}{2}$ and
$\theta=\xi=\tilde{\xi}$. Thus we get a deterministic gate as in
\cite{stator}.

\section{Optimal generation of a single gate}\label{optgate}
Let us return to our original problem. Suppose that Alice and Bob
are interested to implement a gate with a particular value of
$\xi$. Following the scenario presented in Sec. \ref{method} Alice
and Bob will successfully implement the desired gate ($U_{AB}(\xi)$) with
probability $p$ and will fail, i.e. implement the "trash" gate
($U_{AB}(\tilde{\xi})$), with probability $1-p$. In the Fig. \ref{opt} the
probability of success $p$ as a function of $\sin^2\alpha$ is
plotted for various values of $\xi$. Below we have list the important
features of the resulting behavior:

\begin{figure} \epsfxsize=3.5truein
      \centerline{\epsffile{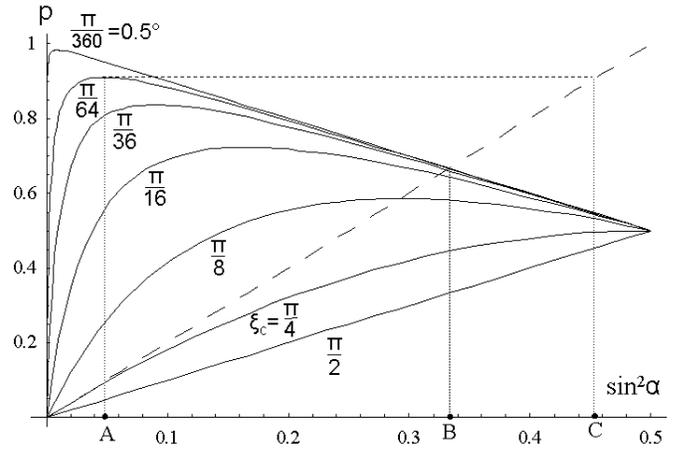}}
  \caption[]{The probability of success $p$ as a function of $\sin^2\alpha$ for various values of $\xi$.
The dashed line shows the probability $2\sin^2\alpha$ given by the
Procrustean method. Points A and C correspond to $\alpha_{opt}$
and $\alpha_{opt}^{Proc}$ for $\xi=\pi/64$. Thus, if the resource
entanglement is in the interval (A,B), then Alice and Bob can
decrease it to optimal A and achieve $p_{max}$ taking an advantage
over the Procrustean method. Point B corresponds to the value
$\sin^2\alpha_c=1/3$, where both methods give same nonzero
probability as $\xi\rightarrow 0$. (The values of $\xi$ presented
on the plot are taken to be proportional to $1/2^n$ for
demonstration only: our method works for all real values of $\xi$
in the appropriate interval.)}
    \label{opt} \end{figure}

\begin{itemize}
\item Above the threshold $\xi_c=\pi/4$ the Procrustean method is
better for all values of $\alpha$. For $\xi<\pi/4$ our method
gives higher probability of success in certain range of $\alpha$.
This range increases as $\xi$ decreases and becomes dominant at
small $\xi$. We note, however, that every gate (\ref{gate}) with
$\xi>\pi/4$ can be decomposed into the gate with $|\xi|<\pi/4$
followed by {\it local} rotations. Thus, the case with $\xi>\pi/4$
is not really interesting from nonlocal point of view.

\item For each fixed value of $\xi$ there is an {\it optimal}
state (\ref{ebit}) with $\alpha=\alpha_{opt}$, where
\beq
\label{pmax} \sin^2\alpha_{opt}=\frac{\sin^2\xi+0.5
\sin2\xi}{\cos2\xi}, \eeq

\noindent which gives the maximum probability of success
$p_{max}$.

\item For a particular value of $\xi$ our method is obviously
better for all $\alpha<\alpha_c$, where
\beq
\sin^2\alpha_c=\frac{2\cos^2\xi-1}{4\cos^2\xi-1} \eeq

\noindent is the point where both methods cross. This value goes
to $1/3$ at the limit $\xi\rightarrow0$. One might think, that for
$\alpha>\alpha_c$ the Procrustean method is always better.
However, this is not the case. Indeed, the Procrustean method
achieves the value equal to $p_{max}$, at some
$\alpha_{opt}^{Proc}>\alpha_c$, where the value of
$\alpha_{opt}^{Proc}$ obeys the condition: \beq
p(\alpha_{opt})=2\sin^2\alpha_{opt}^{Proc}.\eeq

Thus, if Alice and Bob are given by the state (\ref{ebit}) with
$\alpha_{opt}<\alpha<\alpha_{opt}^{Proc}$ then they always can
convert it to the optimal state and achieve $p_{max}$ taking an
advantage over the Procrustean method, which is better only for
$\alpha$ in the range $\alpha_{opt}^{Proc}<\alpha<\pi/4$. In the
following we will show that $\alpha_{opt}^{Proc}$ becomes very
close to $\pi/4$ in the limit of small $\xi$.

 \item For all values of $\xi$ at the limit
$\alpha\ll\xi$ the expression (\ref{p}) can be approximated by a
linear (with respect to $\sin^2\alpha$) dependence \beq
p=\frac{(\sin\alpha)^2}{(\sin\xi)^2},\eeq

\noindent which is consistent with the bound (\ref{bound}), i.e.
this bound is tight at this limit.

\item At the limit $\xi\rightarrow 0$ an additional linear regime
is obtained: if $\xi\ll\alpha$ then $p$ is approximated by \beq
p=1-\sin^2\alpha. \eeq The maximal probability is obtained at
$\alpha_{opt}^2=\xi$: \beq p_{max}=1-\alpha_{opt}^2=1-\xi \eeq


Fig. \ref{small_xi} shows both linear regimes and $p_{max}$ for
$\xi=0.014$ rad.
\begin{figure} \epsfxsize=3.6truein
      \centerline{\epsffile{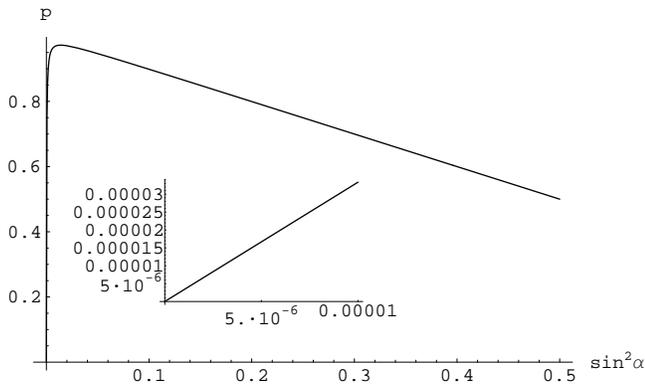}}
        \caption[]{The $p$ vs. $\sin^2\alpha$ dependence for $\xi=0.014$ rad.
        Two linear regimes $p=1-\sin^2\alpha$ and $p=5102\alpha^2$ are presented. The maximal probability
        corresponds to $\alpha_{opt}^2=0.014$ and equals to $p_{max}=0.986$.}
    \label{small_xi} \end{figure}
As we can see, at the limit $\xi\rightarrow 0$ the maximal
probability of success goes to unity as $p_{max}=1-\xi$.
Surprisingly, this probability is obtained for very weakly
entangled optimal state (\ref{ebit}) with $\alpha_{opt}^2=\xi$.

The Procrustean method achieves this probability of success only
for $\sin^2\alpha\geq(1-\xi)/2$, i.e. for $\alpha$ which are very
close to $\pi/4$. In other words, at the limit of small $\xi$ our
method is better that the Procrustean method almost for all
resource states (\ref{ebit}), except for those that are very close
to maximally entangled.

\end{itemize}

From (\ref{xixi}) we can find $\tilde\xi$ of the "trash" gate that
will be generated in the case of failure for every particular
$\xi$ and $\alpha$. We note that each of two gates posses certain
entanglement capability, i.e. ability to generate certain amount
of entanglement. It can be checked explicitly that the
corresponding convex sum of the entanglement capabilities of two
gates does not exceed the entanglement consumed in the process,
i.e. the entanglement possessed by (\ref{ebit}).

\section{Comparison with previous work}
In this section we compare our results with previous work on the
interconvertability of two nonlocal operations \cite{U-U,classes}.
These articles use an isomorphism between states and operations,
in order to generate a transformation between two non-local gates.
In the sense already discussed in the introduction, this method is
indirect:
 $U_1\rightarrow \psi_1 \rightarrow \psi_2\rightarrow U_2$.
First the given non-local gate $U_1$ is applied on a standard
non-entangled state and gives rise to the entangled state
$\psi_1$. Then at an intermediate step, $\psi_1$ is mapped
(distilled) to the desired entangled state $\psi_2$, and finally
the isomorphism is used again to regenerate the desired non-local
gate $U_2$.

The main motivation of \cite{U-U,classes} was searching for
equivalent classes, i.e. the {\it possibility} of such a
transition, so a probability optimization was not an issue in
general - any nonzero probability was good in principle. Thus, the
problem of optimizing the probability for interconversion between
two non-local gates $U_{AB}(\alpha)$ and $U_{AB}(\xi)$ was not
addressed in general, however an answer was given for the special
case of $\xi=\pi/4$. In this case, any $U_{AB}(\alpha)$ can be
deterministically obtained from $U_{AB}(\pi/4)$, while
$U_{AB}(\pi/4)$ can be obtained from $U_{AB}(\alpha)$ with optimal
probability $2(\sin\alpha)^2$, that follows from majorization
condition used on the second step $(\psi_{\alpha}\rightarrow
\psi_{\xi})$ and from the fact that $p[\psi_{\xi}\rightarrow
U_{AB}(\xi)]=1$ in this case. In the more general case,
$\alpha<\xi<\pi/4$, however, the optimal probability
$p[\psi_{\xi}\rightarrow U_{AB}(\xi)]$ is not known. Moreover, the
results of our present paper suggest that using state-operation
isomorphism might be not the most efficient way to address the
problem of efficient non-local gate interconversion, due to the
lose of efficiency at the intermediate indirect
$\psi_{\alpha}\rightarrow \psi_{\xi}\rightarrow U_{\xi}$
transformation. Indeed, for $\alpha<\xi$ our method takes
$\psi_{\alpha}$ directly to $U_{AB}(\xi)$ and achieves the maximal
probability (\ref{bound}). If, however, one goes through
$\psi_{\xi}$ at an intermediate stage then the same value of
probability is multiplied by $p[\psi_{\xi}\rightarrow
U_{AB}(\xi)]$, which is up to date known to be $1/2$ according to
\cite{vidal+cirac} and our present work.

\section{Conclusion}
In this paper we have presented a method for probabilistic
implementation of nonlocal operations (characterized by an
interaction parameter $\xi$) on two qubits using single
nonmaximally entangled state (characterized by an entanglement
monotone $\alpha$) shared by the parties. Our method is
characterized by a direct utilization of a nonmaximally entangled
pair without having to convert it first to a maximally entangled
pair as an intermediate step. We found that if the resource
nonmaximally entanglement is below a certain value, our method has
a higher success probability than the distillation-based (e.g.
Procrustean) methods. This critical value increases with
decreasing of $\xi$, so our method is almost always preferable at
small $\xi$. For very small $\xi$ the probability reaches the
value $p_{max}= 1-\xi$ in contrast to $p=2\xi$ that is given by
the Procrustean method for the same $\alpha$. The explanation of
this efficiency gain seems to be that our method, unlike the
Procrustean method, does not have to create maximal entanglement
at an intermediate stage of the protocol. On the other hand, one
might think that an advantage of the distillation-based methods is
that they do not affect the target state in the case of failure.
We note, however, that our method changes the state in a {\it
known} way.

Interestingly, we found that for higher success probability our
method requires as a resource pairs with less entanglement. For
example, for given $\alpha$ and very small $\xi$ the maximal
probability of success is obtained at $\sin^2\alpha=\xi$. Thus,
when a state with a higher then optimal entanglement is given, the
parties have to reduce the entanglement to the optimal lower
value, before implementing our protocol. This behavior is
consistent with the results of \cite{nlPOVM}.

We have also addressed the fundamental question of the theoretical
bound on the probability of successful implementation of a
nonlocal gate. We found an upper bound for certain range of
parameters and showed that this bound is tight. However, the
general question of the upper bound for the whole range of
parameters remains open.

\begin{acknowledgments}
We would like to thank Sandu Popescu for very useful discussions.
BG acknowledges the European IST-FET project RESQ.  BR
acknowledges the ISF, grant 62/01-1.
\end{acknowledgments}


\begin{thebibliography}{20}
\bibitem{open} This is always the case when $\alpha<\xi$ (see Sec.
\ref{secbound}). However, for $\alpha\geq\xi$ the existence of
deterministic procedure is an open question.
\bibitem{thanks-sandu} As was first pointed out by Sandu Popescu
(privite communication).
\end{thebibliography}
\end{document}